\pdfoutput=1

\documentclass[11pt]{article}

\usepackage[]{acl}
\usepackage{multirow}
\usepackage{booktabs}
\usepackage{makecell}
\usepackage{times}
\usepackage{latexsym}
\usepackage{algorithm}
\usepackage{algorithmic}
\usepackage{arydshln}
\usepackage{amsmath} 
\usepackage{booktabs}
\usepackage{footnote}
\usepackage[T1]{fontenc}

\usepackage[utf8]{inputenc}

\usepackage{microtype}

\usepackage{inconsolata}

\usepackage{graphicx}
\usepackage{xcolor}
\title{How Do Your Code LLMs Perform? Empowering Code Instruction Tuning with High-Quality Data}

\author{Yejie Wang$^1$\thanks{\quad Equal contribution.}, Keqing He$^2$\footnotemark[1], Dayuan Fu$^1$\footnotemark[1], Zhuoma Gongque$^1$, Heyang Xu$^1$\\ \textbf{Yanxu Chen$^1$, Zhexu Wang$^1$, Yujia Fu$^1$, Guanting Dong$^1$, Muxi Diao$^1$}\\ \textbf{Jingang Wang$^2$, Mengdi Zhang$^2$, Xunliang Cai$^2$, Weiran Xu$^1$}\thanks{\quad  Corresponding author.}\\
      $^1$Beijing University of Posts and Telecommunications, Beijing, China\\ $^2$Meituan, Beijing, China \\ 
      \texttt{\{wangyejie,fdy,xuweiran\}@bupt.edu.cn}\\
      \texttt{\{hekeqing,zhangmengdi02,wangjingang02,caixunliang\}@meituan.com}
      }

\begin{document}
\maketitle
\begin{abstract}
Recently, there has been a growing interest in studying how to construct better code instruction tuning data. However, we observe Code models trained with these datasets exhibit high performance on HumanEval but perform worse on other benchmarks such as LiveCodeBench. Upon further investigation, we find that many datasets suffer from severe data leakage. After cleaning up most of the leaked data, some well-known high-quality datasets perform poorly. This discovery reveals a new challenge: identifying which dataset genuinely qualify as high-quality code instruction data. To address this, we propose an efficient code data pruning strategy for selecting good samples. Our approach is based on three dimensions: instruction complexity, response quality, and instruction diversity. Based on our selected data, we present XCoder\footnote{Models and dataset are released in \url{https://github.com/banksy23/XCoder}}, a family of models finetuned from LLaMA3. Our experiments show XCoder achieves new state-of-the-art performance using fewer training data, which verify the effectiveness of our data strategy. Moreover, we perform a comprehensive analysis on the data composition and find existing code datasets have different characteristics according to their construction methods, which provide new insights for future code LLMs.

\end{abstract}

\section{Introduction}

Code pre-trained models have achieved remarkable progress in the era of large language models (LLMs), such as Codex \cite{Chen2021EvaluatingLL}, AlphaCode \cite{Li2022CompetitionlevelCG}, PaLM-Coder \cite{Chowdhery2022PaLMSL} and StarCoder \cite{Li2023StarCoderMT}. Training on large code corpora \cite{Kocetkov2022TheS3} has been shown to enhance the coding capabilities of current LLMs \cite{lozhkov2024starcoder,Rozire2023CodeLO}. In addition to costly pre-training, recent research has garnered increased interest in code instruction tuning and obtains promising results on several code benchmarks \cite{codealpaca,Luo2023WizardCoderEC,codeqwen1.5,Wei2023MagicoderSC,yang2024qwen2,song2024cs,Muennighoff2023OctoPackIT,Wang2024DolphCoderEC}.

\begin{figure}[t]
\centering
\includegraphics[width=0.48\textwidth]{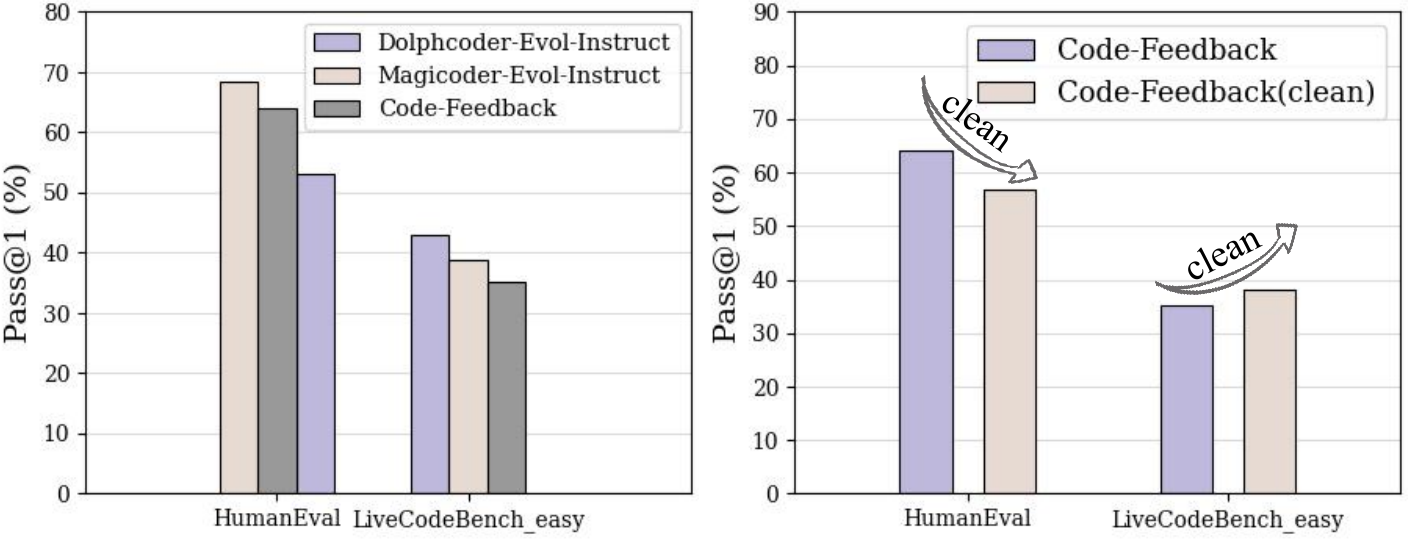}
\caption{The left figure shows performance comparison on different benchmarks and the right displays varying results after data decontamination. Magicoder Evol-Instruct and Code-Feedback may have data leakage on HumanEval.}
\label{fig:intro}
\end{figure}

Differing from the high demand of pre-training for data quantity, instruction tuning aligns existing model abilities towards a desired direction using high-quality but much smaller datasets. To construct code instruction datasets, earlier research predominantly relies on heuristic automation (e.g. distillation from ChatGPT) or manual selection. For example, Code Alpaca \cite{codealpaca} and WizardCoder \cite{Luo2023WizardCoderEC} use distillation signals from ChatGPT via self-instruct and evol-instruct. Other methods such as OctoPack \cite{Muennighoff2023OctoPackIT} and Magicoder \cite{Wei2023MagicoderSC} construct code instructions from pre-training code corpora. Although these code instruction datasets seem excellent on popular code benchmarks like HumanEval\footnote{\url{https://github.com/openai/human-eval}}, we find some of them dramatically drop on another contamination-free benchmark LiveCodeBench \cite{jain2024livecodebench} which continuously collects new problems over time from online contests. As shown in Figure \ref{fig:intro}, Magicoder Evol-Instruct and Code-Feedback \cite{Zheng2024OpenCodeInterpreterIC} achieve top ranks on HumanEval but drop on LiveCodeBench. We perform a further decontamination process and find that several existing code models achieve abnormally high performance on HumanEval because of the potential use of the benchmark or benchmark-similar data. Thus, it remains unclear what good code instruction data is and how these datasets actually work. Besides, all the data come from different pipelines and have no unified principle to ensure good quality. We need to systematically define what constitutes good examples of data for code instruction tuning and establish an effective principle for achieving competitive performance using only highly valuable samples.

In this work, we aim to define the characteristics of good data for code instruction tuning based on a diverse range of existing code datasets. Our goal is to select the most influential samples through a comprehensive and quantitative data assessment measure. Drawing inspiration from \citet{liu2024makes,Ni2024ExploringTM}, we propose a paradigm of data-efficient instruction tuning for code capabilities. Generally, we assume good code samples are complex, of high quality, and diverse. For the complexity aspect, we adopt the evolved complexity scorer to predict the complexity of a given instruction. The scorer is trained on evolved samples via the complexity prompt \cite{Luo2023WizardCoderEC} with ChatGPT. For the aspect of quality, we train a verified model to generate multiple test cases given an (instruction, response) pair and evaluate its quality via the pass rate of the generated test cases. For the aspect of diversity, we select the sample with a large distance to a data pool via instruction embeddings. Combining the three measures, our simple but effective data selection strategy pursues valuable code instruction data and achieves more efficient instruction tuning where fewer training samples yield performance on par with, or even surpassing, models trained on significantly larger datasets. Moreover, we also analyze the composition of our selected data mixture and give suggestions for future code instruction tuning research.

We present XCoder, a family of models finetuned from LLaMA3\footnote{\url{https://LLaMA.meta.com/LLaMA3/}} using our selected code instruction data mixture. Experiments on LiveCodeBench
and HumanEval demonstrate that XCoder is able to outperform or be on par with state-of-the-art code instruction models such as WizardCoder \cite{Luo2023WizardCoderEC}, Magicoder \cite{Wei2023MagicoderSC}, StarCoder2-Instruct\footnote{\url{https://github.com/bigcode-project/starcoder2-self-align}} and OpenCodeInterpreter \cite{Zheng2024OpenCodeInterpreterIC} while using fewer automatically selected data examples. For example, XCoder-8B based on LLaMA3-8B achieves 43.66 LiveCodeBench-Easy and 54.9 HumanEval when trained on only 40K data samples. Besides, our XCoder-70B based on LLaMA3-70B achieves top-tier results compared to the state-of-the-art open-source models.

\section{Deep Dive into Existing Datasets}

\begin{table*}[t]
  \centering
  \resizebox{\textwidth}{!}{%
  \begin{tabular}{lccc}
    \toprule
\textbf{Dataset} & \textbf{Data Size} & \textbf{Instruction Source} & \textbf{Response Source} \\
\hline
    Code-290k-ShareGPT-Vicun \cite{code-sharegpt-vicuna} & 289k & - & -  \\
    CodeExercise-Python-27k~\cite{CodeExercise-Python-27k} & 27k & GPT & GPT   \\
    CodeUp~\cite{CodeUp} & 19k & GPT(Self-Instruct) & GPT   \\
    Glaive-code-assistant-v3~\cite{glaive-code-assistant-v3} & 950k & Glaive & Glaive   \\
    oa\_leet10k~\cite{oa-leet10k} & 23k & - & -   \\
    Code-Alpaca~\cite{codealpaca} & 20k & GPT(Self-Instruct) & GPT   \\
    Codefuse-Evol-Instruct~\cite{liu2023mftcoderboostingcodellms}& 66k & GPT(Evol-Instruct) & GPT   \\
    DolphCoder ~\cite{wang2024dolphcoder} & 79k & GPT(Evol-Instruct) & GPT   \\
    Magicoder-Evol-Instruct ~\cite{Wei2023MagicoderSC} & 110k & GPT(Evol-Instruct) & GPT   \\
    Magicoder-OSS-Instruct ~\cite{Wei2023MagicoderSC} & 75k & GPT(OSS-Instruct) & GPT   \\
    CommitPackFT ~\cite{Muennighoff2023OctoPackIT} & 702k & GitHub & GitHub   \\
    StarCoder2-Self-Align~\cite{sc2-self-align} & 50k & StarCoder2(OSS-Instruct) & StarCoder2   \\
    Leet10k\_alpaca~\cite{leet10k-alpaca} & 10k &- & -   \\
    \bottomrule
  \end{tabular} 
  }

  \caption{Open-source code instruction tuning datasets. Self-Instruct \cite{alpaca} uses LLMs to generate new instructions based on a seed instruction set. Evol-Instruct \cite{xu2023wizardlm,luo2023wizardcoder} use In-Depth Prompts to generate more compelxity instructions. OSS-Instruct \cite{Wei2023MagicoderSC} synthesises diversity instructions through real code snippets.}
  \label{tab:datasets}
\end{table*}

\begin{table*}
  \centering
  \small
    \renewcommand{\arraystretch}{1.2} 
    \setlength{\tabcolsep}{2mm} 
  \begin{tabular}{lllllll}
  \toprule
   \multirow{2}{*}{\textbf{Dataset}} &  \multirow{2}{*}{\textbf{Size}} &  \multirow{2}{*}{\textbf{TLI}}& \multicolumn{2}{c}{\textbf{HumanEval}} & \multicolumn{2}{c}{\textbf{LiveCodeBench}} \\
  \cmidrule{4-7}
 & & & \textbf{\small Base-Pass@1} & \textbf{\small Plus-Pass@1} & \textbf{\small Pass@1} & \textbf{\small Easy-Pass@1} \\
    \hline
\textbf{Codefuse-Evol-Instruct}& 66862 & 8.9 & 61.0 & 53.7 & 13.5 & 34.5 \\
+Clean& 66404 \textcolor{red}{(-0.7\%)} & 4.8 \textcolor{red}{(-4.1)} & 59.1 \textcolor{red}{(-1.9)} & 53.7 \textcolor{red}{(0)} & 12.3 \textcolor{red}{(-1.3)} & 33.1 \textcolor{red}{(-1.4)} \\
\hdashline
\textbf{Magicoder-Evol-Instruct} & 111183 & 43.2 & 68.3 & 64.0 & 15.3 & 38.7  \\
+Clean & 108063  \textcolor{red}{(-2.8\%)} & 4.9 \textcolor{red}{(-38.3)} & 65.9 \textcolor{red}{(-2.4)} & 59.8 \textcolor{red}{(-4.2)} & 13.0 \textcolor{red}{(-2.3)} & 34.5 \textcolor{red}{(-4.2)} \\
\hdashline
\textbf{Code-Feedback} & 66383 & 30.5 & 64.0 & 57.3 & 13.8 & 35.2 \\
+Clean & 64134 \textcolor{red}{(-3.4\%)} & 4.6 \textcolor{red}{(-25.9)} & 56.7 \textcolor{red}{(-7.3)} & 51.8 \textcolor{red}{(-5.5)} & 14.8 \textcolor{red}{(+1.0)} & 38.0 \textcolor{red}{(+2.8)} \\

    \bottomrule
 
  \end{tabular}
  \caption{Comparison of performance across three datasets with data leakage and their cleaned versions on HumanEval and LiveCodeBench. TLI measures the extent of data leakage in the training set on HumanEval. Size and performance changes after cleaning are highlighted in red.}
  \label{tab:data_lekage}
\end{table*}

We present mainstream and open-source Code Instruction Tuning datasets in Table \ref{tab:datasets}. And then we select several influential datasets from these for training and test their performance on HumanEval and LiveCodeBench benchmarks, with the results shown in Table \ref{tab:data_lekage}.

From the results, we observe that different training datasets lead to significant performance differences on HumanEval, but the differences on LiveCodeBench are minimal. This phenomenon leads us to suspect whether the remarkably high performance of some data in HumanEval is due to data leakage. Therefore, we propose the \textbf{T}est \textbf{L}eakage \textbf{I}ndex (TLI) to detect the degree of data leakage for each dataset in the test set.

\paragraph{TLI}
The Test Leakage Indicator is a metric for quantifying the extent of data leakage from a training set to a test set. To compute TLI, n-grams are generated for both datasets, and the overlap between the n-grams of each test sample and those of all training samples is measured. The similarity score \( S(t_i, r_j) \) between a test sample \( t_i \) and a training sample \( r_j \) is calculated as the fraction of common n-grams over the total n-grams in the test samples. For each test sample, the maximum similarity score among all training samples is recorded. The final TLI metric is the average of these maximum similarity scores across all test set. Higher TLI values indicate greater risks of leakage, highlighting significant similarities between the training and test data.

\begin{figure*}[t]
\centering
\includegraphics[width=0.90\textwidth]{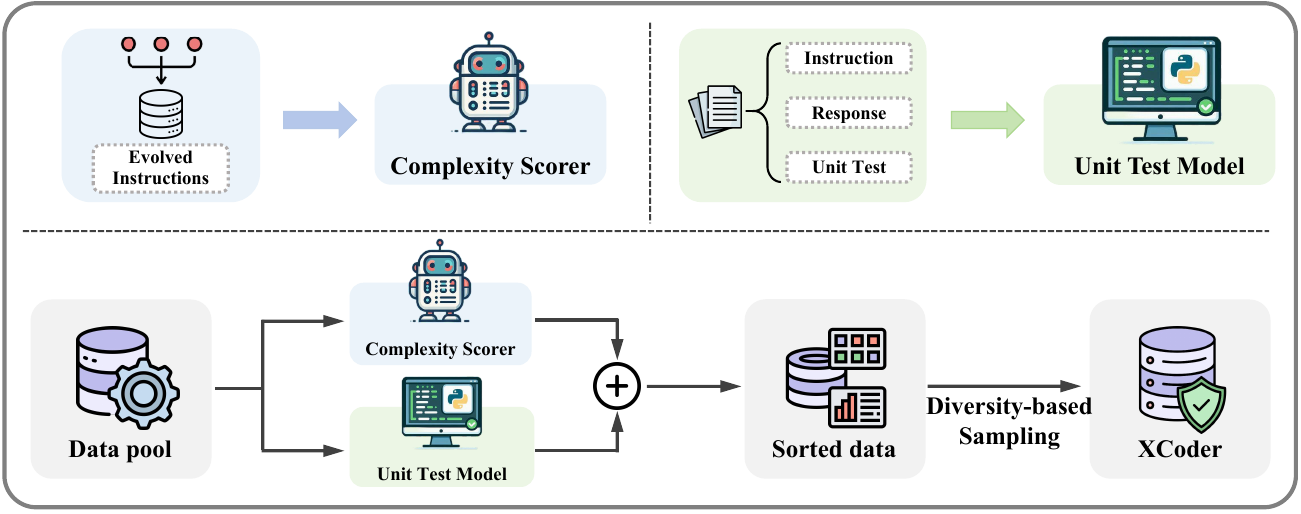}
\caption{Illustration of our data selection approach.
}
\label{fig:main fig}
\end{figure*}

\begin{figure}[t]
    \centering
    \small
    \renewcommand{\arraystretch}{0.8} 
    
    \begin{minipage}{.45\textwidth}

        \begin{algorithm}[H]
        \caption{Data Selection For XCoder}
        \label{alg:training_data_construction}
        \begin{algorithmic}[1]
        \STATE \textbf{Input:} Code Instructing Tuning Data Pool $P = \{ (I_1, R_1), (I_2, R_2), \ldots, (I_N, R_N) \}$, Num of data samples to be selected $Q$, Complexity Scorer $C$, Unit Test Model $U$, Code Interpreter $E$, Hyperparameter $\tau$, Weight $\alpha$. 
        
        \STATE \textbf{Output:} The selected subset $D$
        
        \STATE Initialize Empty Dataset D
        \FOR{$i = 1$ \TO $N$}
            \STATE $c_i \gets C(I_i)$
            \STATE $u_i \gets U(I_i,R_i)$
            \STATE $q_i \gets E(u_i)$
        \ENDFOR

        \FOR{$i = 1$ \TO $N$}
            \STATE $c_i' \gets Normalized(c_i$)
            \STATE $q_i' \gets Normalized(q_i$)
            \STATE $s_i \gets \alpha \times c_i' + (1 - \alpha) \times q_i'$
        \ENDFOR
        
        \STATE $P^* \gets \text{sort}(P, \text{key} = s, \text{reverse} = True)$
        
        \FOR{$k = 1$ \TO $N$}
            \STATE // $distance(I_k, D)$ denotes the distance between $I_k$ and its nearest neighbor in $D$
            \IF{$distance(I_k, D) < \tau$}
                \STATE $D \gets D \cup \{ (I_k, R_k) \}$
            \ENDIF
            \IF{$|D| \geq Q$}
                \STATE \textbf{break}
            \ENDIF
        \ENDFOR

        \end{algorithmic}
        \end{algorithm}
    \end{minipage}
\end{figure}

We calculate the TLI metrics for different datasets on HumanEval, as shown in Table \ref{tab:data_lekage}. More dataset can be viewed in Appendix \ref{appendix:datalekage}. we find that most datasets maintain a TLI of around 5\% on HumanEval, but Codefuse-Evol-Instruct, Magicoder-Evol-Instruct, and Code-Feedback exhibit TLI indices exceeding 30\%. Therefore, we further clean these datasets ensuring that the TLI of all cleaned datasets is controlled at 5\%, and then conduct re-experiments with these datasets. From the result we can observe that the cleaned datasets, after filtering only a small portion, show a significant performance drop on HumanEval, but their performance on LiveCodeBench remains almost unchanged or even slightly improved. For example, after filtering out 3.4\% samples from the Code-Feedback dataset, its performance on the HumanEval Base-Pass@1 metric drops by 7.3\%, but its performance on LiveCodeBench slightly increases. This further substantiates the presence of data leakage. Additionally, we discover numerous cases where the training data are almost identical to the test data in HumanEval, confirming the serious data leakage in these datasets. The leaked cases can be viewed in Appendix \ref{appendix:datalekage}.

\section{What Characteristics Do Good Data Have}

In this section, we first define the characteristics of good data for code instruction tuning and then select the most influential samples via data pruning. Inspired by Deita \cite{liu2024makes}, we select the samples in the Data Pool from three dimensions: instruction complexity, response quality, and instruction diversity. For a data pool $P$, we first use the a complexity score $C$ and Unit Test Model $U$ to calculate the complexity score $c$ and quality score $q$ for each data. Then, we use linearly combine $c'$ and $q'$ to obtain a score $s$ representing complexity and quality. Finally, we sort the data pool $P$ and apply the Diversity-based Sampling to iteratively select samples from the data pool into the final training set $D$, until $D$ reaches the budget size. Our data selection approach is illustrated in the Figure \ref{fig:main fig} and Algorithm \ref{alg:training_data_construction}. The details of Complexity Score, Unit Test Model and Diversity-based Sampling are as follows.

\subsection{Instruction Complexity: Complexity Scorer}
Inspired by Evol Complexity \cite{liu2024makes}, which is a complexity measure based on the evolution algorithm. We use evolved instructions to train our complexity scorer. Specifically, we use self-instruct to obtain a small-scale dataset $Seed = \{ S_1, S_2, \ldots, S_N \}$ as the seed for evolution. Then, we apply the in-depth evolving prompting from WizardCoder for $M$ rounds of evolution. This process results in an instruction set where each seed instruction $s_i$ has $M$ evolved instructions and their corresponding rounds $ \{ (S_i,0), (I_1,1), \ldots, (I_M,M)\}$. We then treat the rounds as a complexity measure and train the complexity scorer to predict the complexity score given the input instruction. In multi-turn dialogues, we score each turn separately and use the sum of them as the final score.

\subsection{Response Quality: Unit Test Model}
We consider the number of test cases passed as a measure of response quality, which, as demonstrated in our experiments in Section \ref{sec:Quality Dimension}, is an effective way to assess code quality for code generation tasks compared to directly scoring the language model.

To obtain test cases for each training sample, we utilize a unit test model that can generate a fully executable unit test program according to the provided instructions and code snippet for testing, which can be formulated as: $T = U(I, R)$, where we denote the instruction as $I$, the code solution as $R$, and the generated unit test as $T$. We collect 6k TACO\cite{li2023taco} data to train the unit test model based on LLaMA3-70B-Base. During application, we prompt the Unit Test Model to generate 12 test cases for each training sample, and execute the unit testing program. The number of passed test cases is considered as the quality score.

We also show some cases output by our unit test model which can be found in Appendix \ref{aPP:B}.

\subsection{Instruction Diversity: Diversity-based Sampling}
We use Diversity-based Sampling method to ensure the diversity of the selected data. The iterative method selects samples $P_i$ one by one from the pool $P$, and when $p_i$ contributes to the diversity of the selected dataset $D$, it is added to $D$. This process continues until the budget $Q$ is reached or all samples $p_i$ in $P$ have been enumerated. Specifically, the benefit of the diversity brought by the newly considered sample $p_i$ can be formulated as an indicator function $F(p_i,D) := distance(p_i, D) < \tau$, which equals 1 only when $F(p_i,D)$ is true, otherwise it is 0. Only when $F(p_i,D)$ equals 1, $p_i$ will be added to $D$. We use the embedding distance between the sample $p_i$ and its nearest neighbor in $D$ to calculate $distance(p_i, D)$. And $\tau$ is a hyperparameter.

\section{Experiments}

\subsection{Benchmarks}
\begin{itemize}
    \item \textbf{HumanEval}: HumanEval \cite{chen2021evaluating} is a widely researched benchmark test for code language models, specifically designed to evaluate the ability of code generation. It includes 164 hand-written programming problems, each problem includes a function signature, docstring, body, and several unit tests, with an average of 7.7 tests per problem. 

    \item \textbf{LiveCodeBench}: LiveCodeBench \cite{jain2024livecodebench} is a comprehensive and pollution-free benchmark for evaluating Large Language Models in code assessment. It updates new problems in real-time from competitions on three competitive platforms (LeetCode, AtCoder, and CodeForces). 
\end{itemize}

\subsection{Implementaion Details}

\begin{table*}[t]
  \centering
  \small
    \renewcommand{\arraystretch}{1.2} 
    \setlength{\tabcolsep}{2mm} 
  \begin{tabular}{l c c c c c c c}
  \toprule
  \multirow{2}{*}{\textbf{Dataset}} & \multirow{2}{*}{\textbf{Size}} & \multicolumn{2}{c}{\textbf{LiveCodeBench}} & \textbf{BigCodeBench} & \multicolumn{2}{c}{\textbf{HumanEval}} \\
  \cmidrule{3-7}
 &  & \textbf{\small Pass@1} & \textbf{\small Easy-Pass@1} &  \textbf{\small Pass@1} & \textbf{\small Base-Pass@1} & \textbf{\small Plus-Pass@1}  \\
    \hline
Code-Alpaca & 20k & 0.0 & 0.0 & 11.9 & 30.5 & 25.6  \\
StarCoder2-Self-Align  & 50k & 9.5 & 24.7 & 14.5 & 37.8 & 34.8   \\
Codefuse-Evol-Instruct* & 66k & 12.3 & 33.1 & 25.4 & 59.1 & 53.7  \\
Magicoder-OSS-Instruct& 75k & 12.8 & 33.8 & 22.0 & 54.3 & 50.0   \\
Magicoder-Evol-Instruct* & 100k &  13.0 & 34.5 & 21.8 & \textbf{65.9} & \textbf{59.8}  \\
Code-Feedback* & 64k & 14.8 & 38.0 & 27.0 & 56.7 & 51.8   \\

\hline
XCoder & 40k & 16.5 & \textbf{43.7} & 27.4 & 54.9 & 50.6   \\
XCoder & 80k & \textbf{16.8} & \textbf{43.7} & \textbf{29.6} & 57.3 & 53.0   \\

\bottomrule
    
  \end{tabular}

  \caption{Comparison of the performance using XCoder data and other mainstream data on HumanEval and LiveCodeBench. All models are trained based on LLaMA3-8B-Base and use greedy decoding. For HumanEval, we report both Base-Pass@1 and Plus-Pass@1 results, where Plus-Pass@1 uses more test cases compared to Base-Pass@1 during evaluation. On LiveCodeBench, we report Pass@1 and Easy-Pass@1 results, with Easy-Pass@1 considering only problems categorized as easy, making it more stable and providing better differentiation than Pass@1. * means that the original dataset may have data leakage, and we perform a n-gram decontamination.}
  \label{Tab:Main Results}

\end{table*}

\paragraph{Data Pools}
To construct the best Code Instruction Tuning dataset, we gathered various available open-source datasets, as detailed in Table 1. This resulted in a collection of 2.5M data samples. However, this amount of data is excessively large. To control the size of the Data Pools, we implemented a straightforward filtering process according to the following rules: Firstly, We include datasets proposed by academic work: Magicoder-OSS-Instruct, Magicoder-Evol-Instruct, and Code-Feedback. We also select the longest 200K samples to add to the Data Pools. Following this, we sort the data by complexity score and add the top 200K highest-scoring samples. Finally, we performed deduplication on the Data Pools, resulting in a final dataset of 336K samples.

\paragraph{Complexity Scorer}
We use ChatGPT to evolve the dataset over 4 iterations on Code-Alpaca as the training set and train on LLaMA3-8B-Instruct with a learning rate of 2e-5 for 1 epoch.

\paragraph{Unit Test Model}
We use 6k TACO data to train our unit test model based on LLaMA3-70B-Base. TACO is a dataset for code generation that each sample contains question, code solutions and test cases. We train the final unit test model using a learning rate of 5e-6 over 3 epochs.

\paragraph{Diversity}
We use LLaMA3-8B-Base to get the instruction embedding. We set $\tau$ to 0.945 which means we consider an example $p_i$ could increase the diversity of selected dataset $D$ when the embedding distance between $p_{i}$ and its nearest neighbor is smaller than 0.945.

\subsection{Main Results}
To validate the effectiveness of XCoder, we conducted experiments on LLaMA3-8B-Base, with the results shown in Table \ref{Tab:Main Results}. From the results we can observe that XCoder achieves the best results on LiveCodeBench and BigCodeBench among other open-source dataset. It also  also achieves the best level performance on HumanEval among the clean datasets. Additionally, we observe that XCoder is highly efficient with samples, achieving superior performance on LiveCodeBench and BigCodeBench with only 40K data compared to baselines. As the data size increases further, XCoder continues to improve on HumanEval and BigCodeBench. We also notice that Magicoder-Evol-Instruct and Codefuse-Evol-Instruct still achieve leading results on HumanEval. The reason may be that the decontamination algorithm cannot completely filter out all leaked data, so some data leakage still exists within these training sets on HumanEval.

\begin{figure*}[t]
\centering
\includegraphics[width=0.85\textwidth]{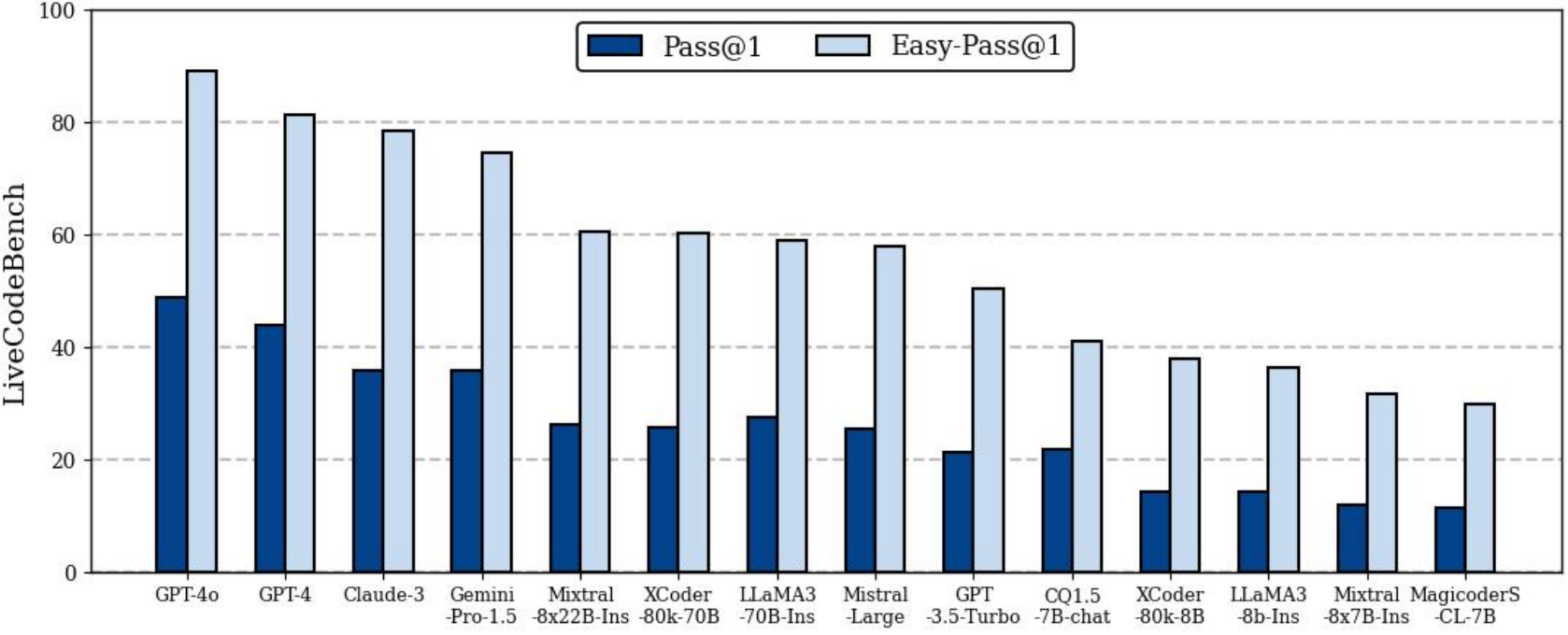}

\caption{Comparison of the performance of XCoder and other mainstream models on LiveCodeBench. Results for other models are sourced from LiveCodeBench Leaderboard~\cite{LiveCode22} For XCoder, we maintain the same settings with other models, where we use 0.2 temperature, sampling 10 solutions for each question. The full name of GPT-4, Glaude-3, Gemini Pro 1.5, GPT-3.5-Turbo, CQ-7B-Chat and MagicoderS-CL-7B are GPT-4o-2024-05-13, GPT-4-Turbo-2024-04-09, Claude-3-opus, Gemini Pro 1.5-May, GPT-3.5-Turbo-0125, CodeQwen15-7B-chat and MagicoderS-CodeLLaMA-7B. We also compare the performance of the model on HumanEval. The complete results can be found in Appendix \ref{appendix_comparion}.}
\label{fig:main results-liv}

\end{figure*}

We also train XCoder-70B based on LLaMA3-70B-Base. Figure \ref{fig:main results-liv} shows that XCoder-70B is one of the best open-source Code LLMs.

\subsection{Analysis}

\subsubsection{Ablation Study}

To validate the effectiveness of each data dimension, we conducted ablation experiments with the results shown in Table \ref{table:module ablation}. As observed across both data sizes, the model's final performance on LiveCodeBench improves with the addition of each dimension, indicating the effectiveness of each dimension.

\subsubsection{Complexity Dimension}

Table \ref{tab:complexity ablation} illustrates the performance of models trained on 40K selected data samples using various complexity measures on LiveCodeBench. Our Complexity Scorer measure exhibits the best performance across all measures, surpassing the Random method by 2.1\% on Pass@1 and by 3.5\% on Easy-Pass@1. The results also indicate that instruction length is a good measure for observing the Code Instruction Tuning data, second only to Complexity Scorer, which contrasts with observations made on general alignment data. Interestingly, perplexity, as an intuitive measure of complexity, performs comparably to the random selection method, consistent with observations by \citet{liu2024makes}.

\begin{table}[t!]
  \centering
  \small
    \renewcommand{\arraystretch}{1.3} 
    \setlength{\tabcolsep}{2mm} 
  \begin{tabular}{lc c c c}
  \toprule
    \multirow{2}{*}{\textbf{Method}} & \multirow{2}{*}{\textbf{Data Size}} & \multicolumn{2}{c}{\textbf{LiveCodeBench}} \\
    \cmidrule{3-4}
 &  &  \textbf{\small Pass@1} & \textbf{\small Easy-Pass@1} \\
  \hline
Random & 40k & 11.5 & 31.0 \\
Complexity & 40k & 13.3 & 34.5 \\
+ Quality & 40k & 15.0 & 39.4 \\
+ Diversity & 40k & \textbf{16.5} & \textbf{43.7} \\
    \hline
Random & 80k & 11.8 & 30.3 \\
Complexity & 80k & 15.0 & 37.3 \\
+ Quality & 80k & \textbf{16.8} & 41.6 \\
+ Diversity & 80k & \textbf{16.8} & \textbf{43.7} \\
    \bottomrule
  \end{tabular}

\caption{We conduct ablation experiments based on LLaMA3-8B-Base with two data sizes to validate the effectiveness of each dimension.}
\label{table:module ablation}
\end{table}

\begin{table}[t!]
  \centering
  \small
    \renewcommand{\arraystretch}{1.3} 
    \setlength{\tabcolsep}{1mm} 

  \begin{tabular}{lc c c c}
  \toprule
    \multirow{2}{*}{\textbf{Measures}} & \multirow{2}{*}{\textbf{Data Size}} & \multicolumn{2}{c}{\textbf{LiveCodeBench}} \\
    \cmidrule{3-4}
 &  & \textbf{\small Pass@1} & \textbf{\small Easy-Pass@1} \\
  \hline
Random & 40k & 11.5 & 31.0 \\
PPL & 40k & 11.8 & 31.0 \\
Length & 40k & 13.0 & 33.1 \\
Complexity Scorer & 40k & \textbf{13.6} & \textbf{34.5} \\
    \bottomrule
  \end{tabular}

  \caption{Comparison of performance on LiveCodeBench using different complexity measurement methods. All models are trained based on LLaMA3-8B-Base and use Greedy decoding. We calculate PPL for each data point using LLaMA3-8B-Base. For the length strategy, we only count the instruction length.}
  \label{tab:complexity ablation}

\end{table}

\subsubsection{Quality Dimension}
\label{sec:Quality Dimension}


\paragraph{Using Unit Test for Ranking}

To validate our Unit Test Model's ability to rank the quality of code, we conducted the following experiment. Specifically, we generate 10 candidate solutions for each question in HumanEval, then use our unit test model to generate test cases for each solution, ranking them based on the number of test cases passed. We select the best one as the final solution. And we use random selection from the candidate solutions as the baseline. The results are shown in Table \ref{tab: quality ablation}. Additionally, we consider another method where using LLMs to output the correctness of the code directly. We choose GPT-4-0409 to do that.
From the results, we observe that compared to random selection, using the unit test model significantly improves the accuracy of the chosen answers, with an increase of nearly 13.6\% in the Base-Pass@1 metric and 10.3\% in the Plus-Pass@1 metric. Notably, the unit test model trained on LLaMA3-70B-Base also outperforms GPT-4, with improvements of around 3\% in both metrics.

From the results, we can observe that using unit tests improves the BoN-Pass@1 metric by approximately 14\%, which is higher than merely using language model judgment. However, we also notice a gap in evaluation accuracy per solution compared to GPT-4. We believe this discrepancy may arise because, for unit tests, a solution must pass all the test cases to be considered correct. Any error in generating a test case can cause the solution to fail. Nevertheless, the effectiveness of unit tests in the Best-of-N metric demonstrates that this approach might be more suitable for ranking the quality of code solutions.

\paragraph{Accuracy of Generated Test Cases}

\begin{table}[ht!]
  \centering
  \small
    \renewcommand{\arraystretch}{1.3} 
    \setlength{\tabcolsep}{1mm} 
  \begin{tabular}{lcc}
  \hline
  \textbf{Method} & \textbf{BoN-Base-Pass@1} & \textbf{BoN-Plus-Pass@1}\\
  \hline
Random & 62.6 & 54.9 \\
GPT-4 & 72.6 & 62.8 \\
Unit Test Model & 76.2 & 65.2 \\
    \hline
  \end{tabular}
\caption{We report the Best-of-N metric on HumanEval. "Random" indicates selecting a solution randomly from the candidates. "GPT-4" involves direct evaluation of each candidate using GPT-4-0409. "Unit Test Model" represents using our unit test model to generate and rank based on test cases passed.}
  \label{tab: quality ablation}
\end{table}

\begin{figure}[t]
\centering
\includegraphics[width=.35\textwidth]{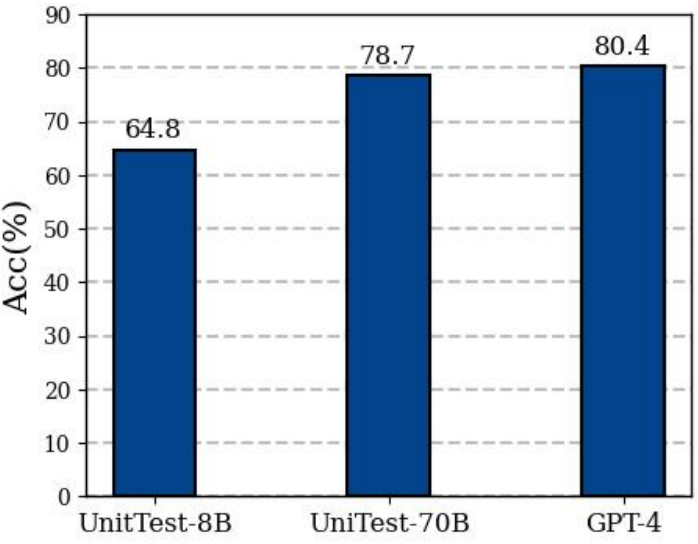}
\caption{Comparison of the accuracy of Unit Test Models trained on different sizes when generating test cases. We also additionally evaluated the ability of GPT-4 to generate test cases.}
\label{fig:unit test acc}
\end{figure}

We also experimented with the impact of different model sizes on the accuracy of the Unit Test Model in generating test cases. Specifically, we instructed the model to generate 10 test cases for the golden solutions in HumanEval, execute them, and count the number of passing test cases. The results are shown in Figure \ref{fig:unit test acc}. Additionally, we evaluated GPT-4's capability in generating test cases.

We observed that increasing the model parameters significantly improves the accuracy of generating test cases, from 64.8\% to 78.7\%. Further, we find that the test case model trained on LLaMA3-70B performs very close to GPT-4 in generating test cases, with a difference of less than 2\%.

\subsubsection{Data Scaling}
To study the impact of our data selection strategy on data scaling efficiency, we conduct experiments using different data budgets. Table \ref{Tab: Data Scaling} shows that XCoder outperforms randomly sampled data across different data sizes. Surprisingly, XCoder achieves performance comparable to using 160K training samples with only 10K samples, and it matches the performance of using the full dataset at 80K samples. This demonstrates the high efficiency of XCoder's data samples and the effectiveness of XCoder in data selection.

\begin{table}
\renewcommand{\arraystretch}{1} 
  \centering
  \small
    \renewcommand{\arraystretch}{1.3} 
    \setlength{\tabcolsep}{2mm} 
  \begin{tabular}{lc c c c}
  \toprule
    \multirow{2}{*}{\textbf{Method}} & \multirow{2}{*}{\textbf{Data Size}} & \multicolumn{2}{c}{\textbf{LiveCodeBench}} \\
    \cmidrule{3-4}
 &  & \textbf{\small Pass@1} & \textbf{\small Easy-Pass@1} \\
  \hline
Random & 10k & 9.8 & 26.1 \\
Random & 40k & 11.5 & 31.0 \\
Random & 80k & 11.8 & 30.3 \\
Random & 160k & 15.0 & 38.8 \\
Random & 320k & 16.8 & 44.4\\
    \hline

XCoder & 10k & 14.5 & 38.0 \\
XCoder & 40k & 16.5 & 43.7 \\
XCoder & 80k & 16.8 & 43.7 \\
XCoder & 160k & \textbf{17.0} & \textbf{44.4} \\
    \bottomrule
  \end{tabular}

  \caption{Comparison of performance on LiveCodeBench with different datasets as the data scales up. We conducted the training on LLaMA3-8B-Base.}
  \label{Tab: Data Scaling}
\end{table}

\subsection{Data Analysis}





In this section, we analyze the data composition of XCoder, reassess the strengths and weaknesses of different data sources, and develop new insights into different data generation methods. 

\paragraph{Complexity:} We sorted all samples according to the Complexity Score and analyzed the source datasets of the top 160K samples. The results are shown in Figure \ref{fig:pie}(a). We observe that the multi-turn Code-Feedback dataset, which includes code refinement data, contributes the largest amount of samples. And OctoPack, which uses real Git commit information as instructions, results in limited instruction complexity and contributes only 0.1\%. However, We also observe that StarCoder2-Self-Align contributes the second largest amount of samples, indicating that, besides Evol-Instruct, converting pre-training data appropriately can also yield complex instructions.

\paragraph{Quality:} Figure 5(b) shows the contribution
of different data sources in the top 160K quality score samples.  We observe that OctoPack, which uses real code data, contributes the most high-quality samples. Moreover, we notice that Magicoder-Evol-Instruct, which used GPT-4 to evolve instructions and generate responses, contributes almost as many high-quality samples as OctoPack. However, Dolphcoder-Evol-Instruct, which used the same Evol-Instruct method but with GPT-3.5 for response generation, only contributes 11.16\% of the samples. And Code-Alpaca, which was generated with text-davinci-003, contributes the fewest high-quality samples, comprising only 2.04\% of the total. We assert that in the Evol-Instruct process, responses generated by more capable models tend to have higher quality. Notably, we observe that StarCoder2-Self-Align contributes a considerable amount, which we think is potentially due to its use of self-synthesized test cases and the rejection of samples that do not execute correctly.

\paragraph{Diversity:} The XCoder method relies on the added samples when calculating the diversity of the samples, meaning it dynamically measures the diversity of the samples and cannot independently calculate diversity scores for each sample. Therefore, we present the composition of the top 160K data before and after applying Diversity-based Sampling method, considering the changes as the impact brought by data diversity. Figure \ref{fig:pie}(c) displays the source statistics of the top 160K samples before using Diversity-based Sampling, while Figure \ref{fig:pie}(d) illustrates the composition of the data after applying Diversity-based Sampling for the top 160K data. We find that the most notable change is that, after applying the Diversity-based Sampling method, OctoPack jumps from having the lowest contribution to the second highest. We believe this phenomenon may be due to OctoPack directly gathering instructions from the real world, thus possessing better diversity.

Overall, we find that in terms of complexity: data with more rounds has longer context and higher complexity. Additionally, Evol-Instruct is an effective method for improving instruction complexity. In terms of quality: Code LLMs that deliver accurate responses. Data with added test case feedback verification during data synthesis tends to have higher quality. Furthermore, using a stronger model to synthesize data is a simpler, more direct, but effective approach. In terms of diversity: We find that directly sampling from the real world and transforming it results in instructions with better diversity compared to other methods that only expand instructions using fixed seeds.

\section{Related Work}
\paragraph{Code Instruction Tuning.} Code instruction tuning is a necessary step for models to accurately understand human instructions and generate relevant code responses. 
\citet{xu2023wizardlm} apply the Evol-Instruct method~\cite{xu2023wizardlm} to Code-Alpaca\cite{codealpaca} dataset and obtain a instruction dataset with high complexity. \citet{Muennighoff2023OctoPackIT}\ take git commits as natural instruction data. They collect 4TB git commits across 350 programming language. 
\citet{wang2024dolphcoder} propose Diverse Instruction Tuning and Multi-Objective Tuning to train Dolphcoder, which proves that more diverse code solutions and code evaluation instruction data are also beneficial for code generation tasks. 
Considering Evol-Instruct depends on a seed instruction data which is less diversity, \citet{Wei2023MagicoderSC} proposes OSS-Instruct, which leverages open-source code cnippets to generate high-diversity instructions. They also propose Magicoder-Evol-Instruct dataset and train Magicoder-S, which is the first 7B model to exceed 70\% on HumanEval Pass@1. However, we find this dataset suffers from serious data contamination~\citep{dong2024abilitieslargelanguagemodels,xu2024benchmarkingbenchmarkleakagelarge}. 
Motivated by various works with execution feedback~\citep{cao2024towards,le2022coderl,chen2023teaching,qiao2023making,qiao2024we,dong2024self}, OpenCodeInterpreter~\cite{Zheng2024OpenCodeInterpreterIC} and AutoCoder~\cite{lei2024autocoder} leverages GPT-4 and Code Interpreter as code feedback to generate multi-turn instruction data which instruct model to refine incorrect code snippets acrroding to feedback information.

\paragraph{Data Selection for Instruction Tuning.} While instruction fine-tuning primarily relies on a large volume of data, research such as LIMA~\citep{zhou2024lima} indicates that data quality is more critical than quantity.
\citet{li2024quantityqualityboostingllm} proposes a novel metric \textbf{I}nstruction \textbf{F}ollowing \textbf{D}ifficulty(IFD) to assess the challenge of responding to specific instructions.
\citet{li2024oneshotlearninginstructiondata} harnesses the disparity between one-shot and zero-shot scores to calculate a definitive 'gold score' for each instruction. 
\citet{kung-etal-2023-active} present Active Instruction Tuning, which introduces the concept of Prompt Uncertainty. Tasks that exhibit higher Prompt Uncertainty are prioritized for instruction tuning.
Furthermore, \citet{lu2023instag} introduce an automated instruction tagging method (INSTAG), which employs ChatGPT to generate detailed, open-ended labels for instructions. It starts by sorting instructions in descending order of label count and then iteratively adds instructions to a subset based on the uniqueness of their labels.
Deita~\citep{liu2024makes} integrates a multifaceted approach for selecting instruction data, focusing on complexity, quality, and diversity. Utilizing the WizardLM technique, ChatGPT is employed to augment instructions, which are then evaluated for both complexity and quality by specially trained scorers. 



\section{Conclusion And Future Work}

Code LLMs have raised great interest in current LLM research and plenty of code instruction datasets are proposed over time. However, although many of them claim that good results are achieved on the popular benchmark HumanEval, we find several datasets may have data leakage by using benchmark samples as seed data in self-instruct or evolve-instruct. In this paper, we aim to identify which dataset genuinely qualifies as high-quality code instruction data and propose an efficient code data selection strategy for selecting valuable samples. Based on three dimensions of assessing data, we present XCoder, a family of models finetuned from LLaMA3 on our selected dataset. XCoder achieves superior performance than the SOTA baselines using fewer training samples. From the composition of our selected data mixture, we find existing code datasets have different characteristics corresponding to their construction methods, which provide new insights for developing better code LLMs.

\section{Limitation}
Our limitations are two-fold: (1) We only explore our method on the LLaMA3-Base model. More experiments on different model bases are needed to confirm our conclusions. (2) We only focus on the Code Generation task, and in the future, we need to incorporate data containing more tasks.

\section{Broader Impacts}
Similar to the other LLMs, our XCoder could
also generate unethical, harmful, or misleading information, which is not considered in our work.
Future research to address the ethical and societal
implications is needed. XCoder is also susceptible to hallucination in ungrounded generation
use cases due to its smaller size. This model is
solely designed for research settings, and its testing
has only been carried out in such environments.
It should not be used in downstream applications,
as additional analysis is needed to assess potential
harm or bias in the proposed application



\bibliography{custom}

\clearpage
\appendix

\section{Other Implementation Details}
\label{appendix:details}
\paragraph{Training Details:} 
We trained on LLaMA3-8B-Base and LLaMA3-70B-Base. For the 8B model, we train with a learning rate of 2e-5, while for the 70B model, we use a learning rate of 5e-6. All models are trained for 2 epochs. The batch size during training varies according to the dataset size: for datasets with fewer than 40K samples, the batch size is set to 256; for datasets between 40K and 80K samples, the batch size is set to 512; for datasets between 80K and 160K samples, the batch size is set to 1024; and for datasets larger than 160K samples, the batch size is set to 2048.

\begin{figure*}[t]
\centering
\includegraphics[width=\textwidth]{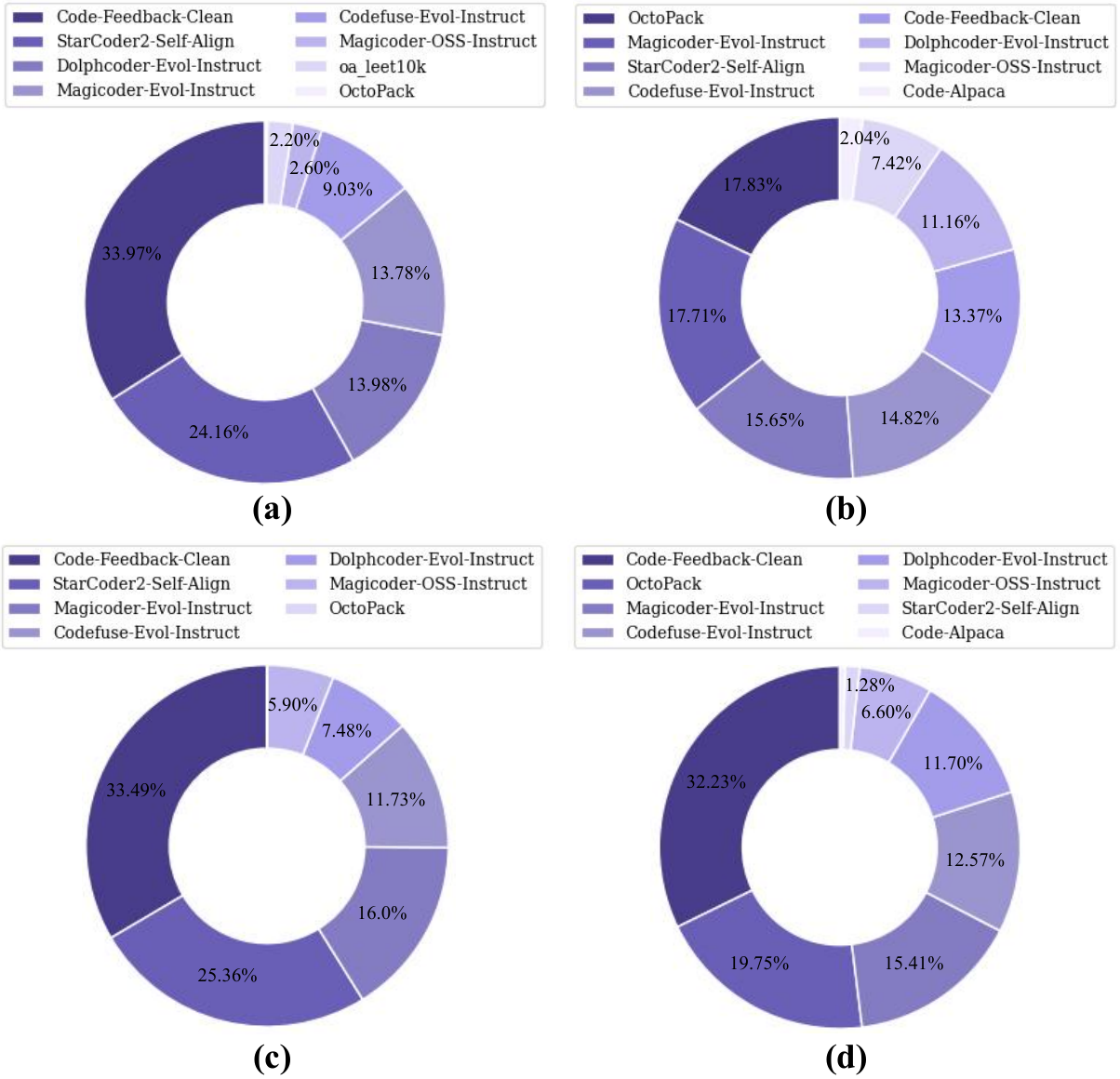}
\caption{ The contribution ratio of different data sources to XCoder, with (a) representing the source of the 160K samples with the highest complexity, (b) representing the 160K samples with the highest quality, and (c) and (d) reflecting which dataset has better diversity.
}
\label{fig:pie}
\end{figure*}

\section{Case Study on Data Leakage}
\label{appendix:datalekage}
\label{sec:Casestudy}
\begin{figure*}[t]
    \centering
    \resizebox{\textwidth}{!}{
    \includegraphics{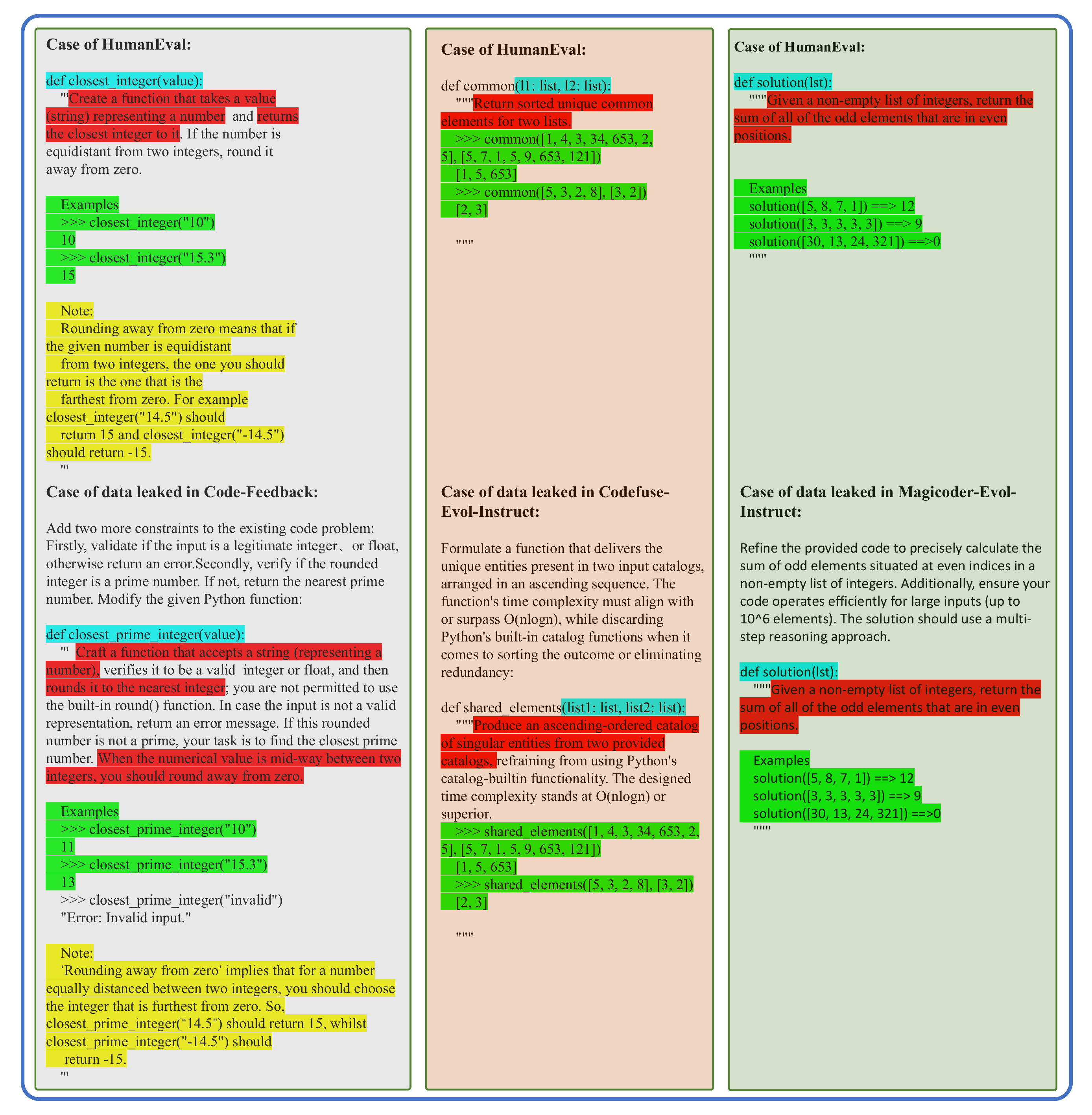}
    }
    \caption{Examples of data leakage.}
    \label{appfig:leakagecase}
\end{figure*}

We show examples of data leakage in Codefuse-Evol-Instruct, Magicoder-Evol-Instruct and Code-Feed-back in Figure \ref{appfig:leakagecase}.
The statistical information of data leakage can be seen in Table \ref{tab:leakage1}.

\begin{table*}
  \centering
  \resizebox{1\textwidth}{!}{
  \begin{tabular}{lllllll}
  \toprule
  \multirow{2}{*}{\textbf{Dataset}} &  \multirow{2}{*}{\textbf{Size}} &  \multirow{2}{*}{\textbf{TLI}}& \multicolumn{2}{c}{\textbf{HumanEval}} & \multicolumn{2}{c}{\textbf{LiveCodeBench}} \\
  \cmidrule{4-7}
 & & & \textbf{\small Base-Pass@1} & \textbf{\small Plus-Pass@1} & \textbf{\small Pass@1} & \textbf{\small Easy-Pass@1} \\
    \hline
\textbf{Code-Alpaca} & 20022 & 3.4 & 30.5 & 25.6 & 0.0 & 0.0 \\
\textbf{StarCoder2-Self-Align} & 50661 & 4.7 & 37.8 & 34.8 & 9.5 & 24.7 \\
\textbf{Magicoder-OSS-Instruct} & 75197 & 4.5 & 54.3 & 50.0 & 12.8 & 33.8 \\
\textbf{Codefuse-Evol-Instruct} & 66862 & 8.9 & 61.0 & 53.7 & 13.5 & 34.5 \\
\textbf{Magicoder-Evol-Instruct} & 111183 & 43.2 & 68.3 & 64.0 & 15.3 & 38.7 \\
\textbf{Code-Feedback} & 66383 & 30.5 & 64.0 & 57.3 & 13.8 & 35.2 \\

\textbf{Codefuse-Evol-Instruct-clean} & 66404 \textcolor{red}{(-0.7\%)} & 4.8 \textcolor{red}{(-4.1)} & 59.1 \textcolor{red}{(-1.9)} & 53.7 \textcolor{red}{(0)} & 12.3 \textcolor{red}{(-1.3)} & 33.1 \textcolor{red}{(-1.4)} \\

\textbf{Magicoder-Evol-Instruct-clean} & 108063  \textcolor{red}{(-2.8\%)} & 4.9 \textcolor{red}{(-4.0)} & 65.9 \textcolor{red}{(-2.4)} & 59.8 \textcolor{red}{(-4.2)} & 13.0 \textcolor{red}{(-2.3)} & 34.5 \textcolor{red}{(-4.2)} \\
\textbf{Code-Feedback-clean} & 64134 \textcolor{red}{(-3.4\%)} & 4.6 \textcolor{red}{(-25.9)} & 56.7 \textcolor{red}{(-7.3)} & 51.8 \textcolor{red}{(-5.5)} & 14.8 \textcolor{red}{(+1.0)} & 38.0 \textcolor{red}{(+2.8)} \\
    \bottomrule
    
  \end{tabular}
}
  \caption{Data Leakage Statistics on HumanEval}
\label{tab:leakage1}
\end{table*}

\begin{figure*}[t]
    \centering
\includegraphics[width=\textwidth]{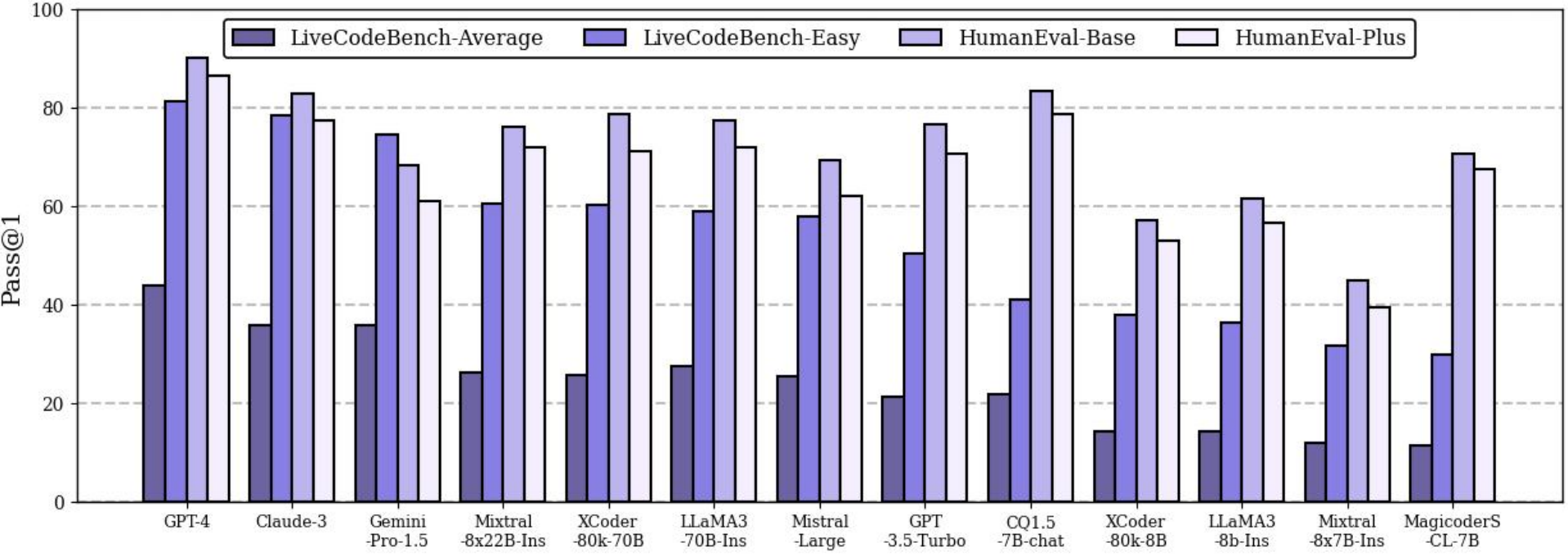}
\caption{Comparison of the performance of XCoder and other mainstream models on HumanEval and LiveCodeBench. Results for other models are sourced from Eval Plus Leaderboard and LiveCodeBench Leaderboard. For XCoder, we maintain the same settings with other models, where for HumanEval we use a greedy decoding strategy and for LiveCodeBench we use 0.2 temperature, sampling 10 solutions for each question. The
full name of GPT-4, Glaude-3, Gemini Pro 1.5, GPT-3.5-Turbo ,CQ-7B-Chat  and MagicoderS-CL-7B are GPT-4o-2024-05-13, GPT-4-Turbo-2024-04-09, Claude-3-opus, Gemini Pro 1.5-May, GPT-3.5-Turbo-0125, CodeQwen15-7B-chat and MagicoderS-CodeLLaMA-7B.
}
\label{fig:main results}
\end{figure*}

\section{Example of input and output for unit test model}
\label{aPP:B}
\begin{figure*}[t]
    \centering
\includegraphics[width=\textwidth]{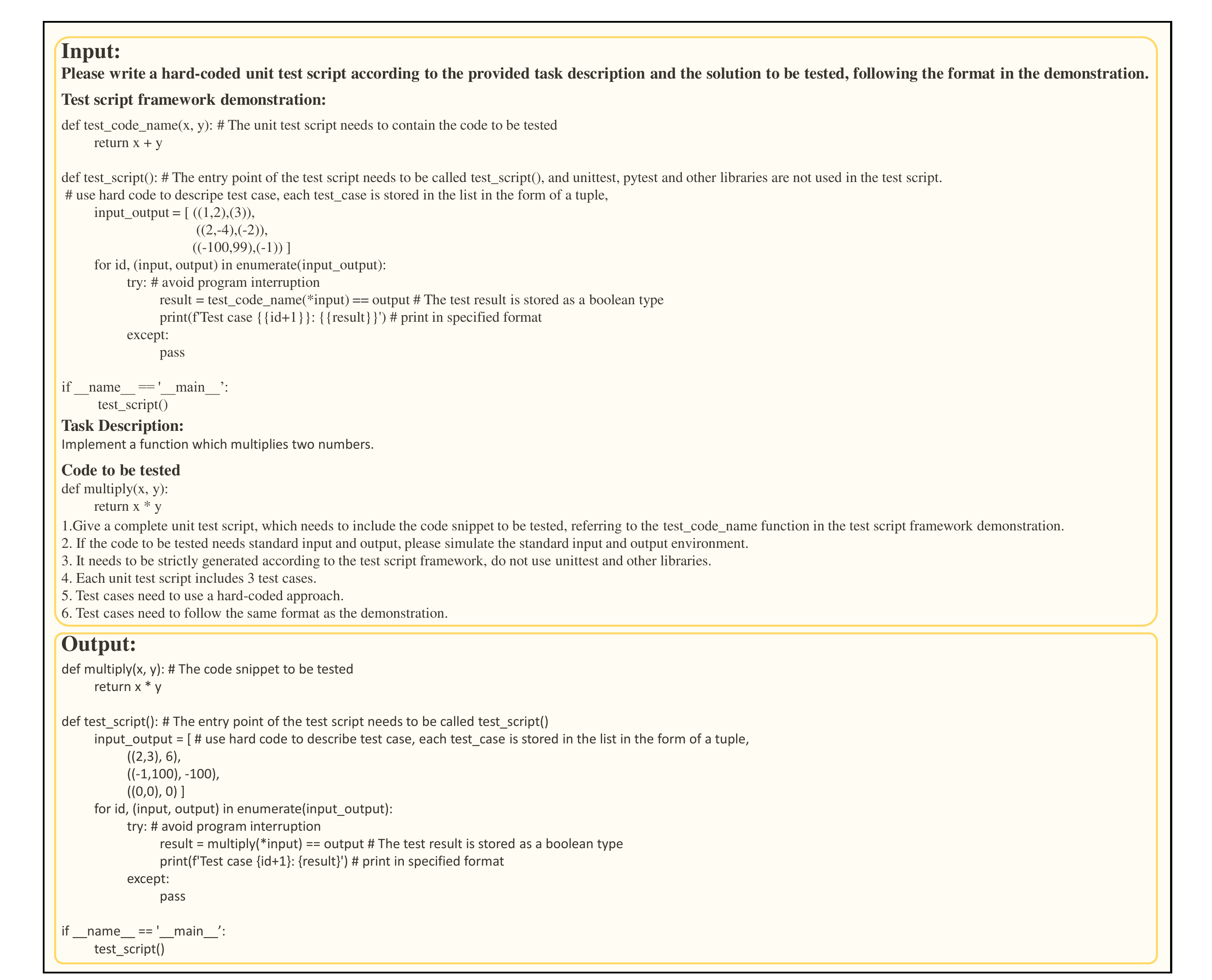}
\caption{Input and output case of unit test model.}
\label{app:unit test case}
\end{figure*}

We present an input and output case of unit test model in Figure \ref{app:unit test case}.

\section{Comparion of XCoder and other mainstream models}
\label{appendix_comparion}
We Compare the performance of XCoder and other mainstream models on HumanEval and LiveCodeBench. The result is shown on Figure \ref{fig:main results} and Table \ref{Tab:Main Results2}

\begin{table*}[t]
  \centering
  \small
    \renewcommand{\arraystretch}{1.2} 
    \setlength{\tabcolsep}{2mm} 
  \begin{tabular}{l c c c c c c c}
  \toprule
  \multirow{2}{*}{\textbf{Model}} & \multirow{2}{*} & \multicolumn{2}{c}{\textbf{LiveCodeBench}} & \multicolumn{2}{c}{\textbf{HumanEval}} \\
  \cmidrule{3-7}
 &  & \textbf{\small Pass@1} & \textbf{\small Easy-Pass@1} & \textbf{\small Base-Pass@1} & \textbf{\small Plus-Pass@1}  \\
    \hline
GPT-4O &  & 48.8 & 89.2 & - & -
\\
GPT-4 &  & 43.9 & 81.5 & 90.2 & 86.6
\\
Claude-3 &  & 35.9 & 78.5 & 82.9 & 77.4 \\
Gemini-Pro-1.5  &  & 35.9 & 74.7 & 68.3 & 61.0 \\
Mixtral-8x22B-Ins &  & 26.4 & 60.7 & 76.2 & 72.0 \\
LLaMA3-70B-Ins & &  27.6 & 59.1 & 77.4 & 72.0 \\
Mistral-Large & & 25.4 & 58.1 & 69.5 & 62.2 \\
GPT-3.5-Turbo & & 21.4 & 50.4 & 76.8 & 70.7 \\
CQ1.5-7B-chat & & 21.8 & 41.0 & 83.5 & 78.7 \\
LLaMA3-8b-Ins & & 14.3 & 36.4 & 61.6 & 56.7 \\
Mixtral-8x7B-Ins & & 12.1 & 31.8 & 45.1 & 39.6 \\
MagicoderS-CL-7B & & 11.4 & 29.9 & 70.7 & 67.7 \\
\hline
XCoder-80k-8B & & 14.4 & 38.1 & 57.3 & 53.0 \\
XCoder-80k-70B & & 25.9 & 60.4 & 78.7 & 71.3 \\

\bottomrule
    
  \end{tabular}

  \caption{Comparison of the performance of XCoder and other mainstream models on HumanEval and LiveCodeBench. Results for other models are sourced from Eval Plus Leaderboard~\cite{EvalPlus12} and LiveCodeBench Leaderboard. For XCoder, we maintain the same settings with other models, where for HumanEval we use a greedy decoding strategy and for LiveCodeBench we use 0.2 temperature, sampling 10 solutions for each question. The full name of GPT-4, Glaude-3, Gemini Pro 1.5, GPT-3.5-Turbo ,CQ-7B-Chat  and MagicoderS-CL-7B are GPT-4o-2024-05-13, GPT-4-Turbo-2024-04-09, Claude-3-opus, Gemini Pro 1.5-May, GPT-3.5-Turbo-0125, CodeQwen15-7B-chat and MagicoderS-CodeLLaMA-7B.}
  \label{Tab:Main Results2}

\end{table*}

\end{document}